\documentstyle[preprint,aps]{revtex}
\begin{document}
\input epsf
\title{
\hspace{4.in} {\normalsize IU/NTC 96-04} \\ 
\hspace{4.in} {\normalsize ADP-96-14/T217} \\ 
\vspace{0.4cm}
Probing Charge-Symmetry-Violating Quark
Distributions in Semi-Inclusive Leptoproduction of
Hadrons}

\author{J. T. Londergan, Alex Pang}
\address{
Dept.\ of Physics and Nuclear Theory Center, Indiana University,
Bloomington, IN 47404
}
\vspace{0.4cm}
\author{A.W. Thomas}
\address{
Dept.\ of Physics and Mathematical Physics and Institute for Theoretical
Physics, 
University of Adelaide, 
Adelaide, S.A., 5005, Australia
}
\date{\today}
\maketitle
\begin{abstract}
Recent experiments by the HERMES group at HERA are 
measuring semi-inclusive electroproduction of pions
from deuterium.  We point out that by comparing 
the production of $\pi^+$
and $\pi^-$ from an isoscalar target, it is 
possible, in principle, to measure charge symmetry
violation in the valence quark distributions of the nucleons.  
It is also possible in the same experiments to obtain an
independent measurement of the quark 
fragmentation functions.  We
review the information which can be deduced from such
experiments and show the ``signature'' for charge symmetry
violation in such experiments. Finally, we predict the
magnitude of the
charge symmetry violation, from both the valence quark distributions and
the pion fragmentation function, which might be expected in
these experiments.  
\end{abstract}

\section{Introduction}

It has long been been recognized that 
charge symmetry or isospin symmetry is respected in the
strong interaction to a high degree of precision.  Charge
symmetry violation [CSV] in nuclear physics is generally good
to within about 1\% of the strong amplitude \cite{nefkens,miller}.  
Therefore, in most analyses of strong interactions, it 
is reasonable to assume the validity of charge symmetry.  

Charge symmetry has always been assumed in defining 
quark distribution functions.  It is so common in
quark/parton phenomenology that 
it is frequently not even mentioned as an assumption.  
Charge symmetry reduces by a factor
of two the number of independent quark distributions one
must define, and until recently there has been no
compelling reason to suggest charge symmetry violation.  
On the other hand, there were no precise tests
of charge symmetry in parton distributions, either.  

At the nucleon level, charge symmetry appears as the
symmetry under interchange of protons with neutrons.  
At the quark level, charge symmetry involves interchange
of up and down quarks.  
Recently, the question of charge symmetry at the
quark level has become of great interest.  At the
present time, we have a good quantitative knowledge
of the valence and sea quark distributions in
the nucleon.  Furthermore, it is now possible to
perform experiments whose precision, and range in
$x$, is sufficient to probe ``small'' components
of quark distributions, and hence to test 
quantities like flavor symmetry of nucleon
sea distributions, and charge symmetry of parton
distributions, which were not accessible to 
direct experimental test in the past.

Interest in this general field has been
sparked by the experimental evidence of violation of the 
Gottfried sum rule \cite{GSR}
reported by the New Muon Collaboration (NMC) \cite{NMC}.  
Although this has widely been quoted as evidence
of SU(2) flavor symmetry violation [FSV] in the proton sea 
\cite{PRS}-\cite{KL} (i.e. $\bar{u}^p(x) \not = \bar{d}^p(x)$),  
an alternative explanation of that result could be
from charge symmetry breaking in the proton sea, 
as has been pointed out in Refs. \cite{Ma92,Ma93} (it could
also arise from a linear combination of CSV 
and FSV effects in the nucleon sea).

Initial theoretical investigations of charge symmetry
in valence quark distributions have been carried
out by Sather \cite{Sather}, and by Rodionov {\it et al.} 
\cite{Rodionov}. Following the approach of Ref.\cite{SIG},
they calculated valence quark distributions 
for the proton and neutron in a quark model 
and in the MIT bag model, respectively.  In both models the ``majority''
quark distributions (i.e., $u^p(x)$ and $d^n(x)$) satisfied
charge symmetry to within about 1\%, while the ``minority''
quark distributions ($d^p(x)$ and $u^n(x)$) were predicted to violate
charge symmetry by 5\% or more at large $x$.  
More recently, Londergan {\it et al.} \cite{londergan} proposed that
pion-induced Drell-Yan processes could be used to probe
these CSV effects in nucleon parton distributions.

In this note, we point out that semi-inclusive pion production, 
from lepton deep inelastic scattering on nuclear targets, 
could also be a sensitive probe of CSV effects in nucleon
valence distributions.  Several authors have previously 
discussed the possibility of using semi-inclusive
scattering processes to address various physics issues 
\cite{LMS,Frankfurt,LuMa}.
In fact, the HERMES collaboration at HERA \cite{HERA} 
is currently taking experimental data on semi-inclusive 
pion production from hydrogen and deuterium.  

The organization of our paper is as follows.  In Sec.\ II, we
review the formalism for semi-inclusive hadron production
at high energies.  We show that for an isoscalar nuclear
target, one can test charge symmetry violation in both
the valence quark distribution functions, and in the 
fragmentation function.  To first order in these small
effects, there is an experimental combination for which these two 
quantities separate.  In Sec.\ III, we derive  predictions
for ratios of $\pi^+$ and $\pi^-$ semi-inclusive 
electroproduction from deuterium.  We show both the
qualitative and quantitative ``signatures'' for charge symmetry
violation in these reactions.  
We present our conclusions and suggestions for
future experimental work in Sec.\ IV.  

\section{Charge Symmetry Tests in Semi-Inclusive Hadron
Electroproduction}

\subsection{General Formulas for Semi-Inclusive
Reactions}

In the quark/parton model, the semi-inclusive production of 
hadrons in deep inelastic lepton scattering from a nucleon 
is given by
\begin{equation}
{1\over \sigma_N(x)}{d\sigma^h_N(x,z)\over dz}=
{\sum_i e^2_i q^N_i(x) D^h_i(z)\over \sum_i e^2_i q^N_i(x)}. 
\label{cs}
\end{equation}
In Eq.\ (\ref{cs}), $q^N_i(x)$ is the distribution function for
quarks of flavor $i$, and charge $e_i$, in the hadron $N$ as a function of
Bjorken $x$.  $D^h_i(z)$ is the fragmentation function
for a quark of flavor $i$ into hadron $h$. The fragmentation
function depends on the quark longitudinal momentum 
fraction $z=E_h/\nu$, where $E_h$ and $\nu$ are the energy of 
the hadron and the virtual photon respectively.
We write the numerator in Eq.\ (\ref{cs}) as $N^{Nh}\equiv
\sum_i e^2_i q^N_i(x)D^h_i(z)$, so the quantity $N^{Nh}$
represents the yield of hadron $h$ per scattering from 
nucleon $N$, as a function of $z$ and $x$.  
In terms of these quantities, 
semi-inclusive production of a charged hadron from
a proton can be described by 
\begin{eqnarray}
N^{p\pm}(x,z) &=& {4\over 9}u^p(x) D^{\pm}_u(z) 
  +{4\over 9}\bar{u}^p(x) D^{\pm}_{\bar u}(z)
  +{1\over 9} d^p(x) D^{\pm}_d(z) \nonumber \\
  &+& {1\over 9}\bar{d}^p(x) D^{\pm}_{\bar d}(z)
  +{1\over 9}s^p(x)D^{\pm}_s(z) 
  +{1\over 9}\bar{s}^p(x) D^{\pm}_{\bar s}(z) ,
\label{genfor}
\end{eqnarray}
where $D^{\pm}_i(z)$ are the fragmentation functions for a quark 
(or antiquark) of flavor $i$ into positively
or negatively charged hadrons.

Charge conjugation invariance implies that 
$D^{\pm}_u=D^{\mp}_{\bar u}$ and $D^{\pm}_d=D^{\mp}_{\bar d}$.
By making the additional assumption of charge symmetry, 
the fragmentation functions
for pions from quarks will obey the relations
\begin{eqnarray}
D_d^{\pi^-}(z) &=& D_u^{\pi^+}(z) \nonumber \\
D_d^{\pi^+}(z) &=& D_u^{\pi^-}(z) 
\end{eqnarray}
Later, we will consider the possibility of charge symmetry
violation in the fragmentation functions.  

We want to find an experimental signature for charge symmetry
violation in valence quark distributions of the nucleon.  
We will therefore derive expressions for $\pi^+$ and $\pi^-$ 
electroproduction on an isoscalar nucleus. In this paper 
our expressions specifically refer to a
deuteron target, although our results can be extended to
any isoscalar target.   

\subsection{Experimental Extraction of Fragmentation
Functions}

Levelt {\it et al.} \cite{LMS} 
derived a useful expression by which
the ratio of fragmentation functions can be extracted 
from leptoproduction of charged pions on protons and
neutrons.  
They proposed measuring the quantity
\begin{eqnarray}
R(x,z) &=& {[ N^{p\pi^+}(x,z)- N^{n\pi^+}(x,z)
   +  N^{p\pi^-}(x,z)-  N^{n\pi^-}(x,z)] \over 
    [ N^{p\pi^+}(x,z)- N^{n\pi^+}(x,z)
   -  N^{p\pi^-}(x,z)+  N^{n\pi^-}(x,z)] } 
\end{eqnarray}
In the quark/parton model, $R(x,z)$ has the form
\begin{eqnarray}
  R(x,z) &=& {3\over 5}\,\left({\tau(x) - \bar{\tau}(x) \over
  \tau(x) + \bar{\tau}(x)}\right) \,\left( {D_u^+(z) + D_u^-(z)\over 
  D_u^+(z) - D_u^-(z)}\right)  , 
\label{lmstau}
\end{eqnarray}
with the definitions
\begin{eqnarray}
 \tau(x) &=& u^p(x) - d^p(x) \nonumber \\
 \bar{\tau}(x) &=& -\bar{u}^p(x) + \bar{d}^p(x) .
\end{eqnarray}

Integrating both the numerator and denominator of $R(x,z)$
over $x$ gave the result      
\begin{eqnarray}
Q(z) &=& \int_0^1\,dx\,[ N^{p\pi^+}(x,z)- N^{n\pi^+}(x,z)
   +  N^{p\pi^-}(x,z)-  N^{n\pi^-}(x,z)]/ \nonumber \\
   &(& \int_0^1\,dx\, [ N^{p\pi^+}(x,z)- N^{n\pi^+}(x,z)
   -  N^{p\pi^-}(x,z)+  N^{n\pi^-}(x,z)] ) \nonumber \\
  &=& {9\over 5}\, S_G\, \left( {D_u^+(z) + D_u^-(z)\over 
  D_u^+(z) - D_u^-(z)} \right)
\label{lmsrat}
\end{eqnarray}
In Eq.\ (\ref{lmsrat}), $S_G$ is the experimental 
value for the Gottfried Sum Rule \cite{NMC}.      

The quantity $Q(z)$ is a function of the ratio of the 
``favored''
and ``unfavored'' fragmentation functions, $D_u^+(z)$
and $D_u^-(z)$, respectively.  They are so named
because production of a positively (negatively) 
charged hadron will preferentially occur from the up 
(down) quark in the nucleon.  The fragmentation
functions have been 
extracted by the EMC
group \cite{emcfrag,emcfrag89}, but an independent measurement
of this can be obtained in pion leptoproduction. 
Feynman and Field \cite{feyfie} suggested a form for
the ratio
\begin{eqnarray}
\Delta(z) &\equiv& {D_u^-(z)\over D_u^+(z)} 
 = {1- z\over 1+z}
\label{ffrat}
\end{eqnarray}
If Eq.\ (\ref{ffrat}) is correct, then measurement of
Eq. (\ref{lmsrat}) would give
\begin{equation}
Q(z) = {9\over 5z}\,S_G 
\label{Rlms}
\end{equation}

The quantity $Q(z)$ of Eq. (\ref{lmsrat}) requires charged 
hadron production from both protons and
neutrons.  For charged pion leptoproduction on an
isoscalar target (e.g., deuterium), a useful quantity
is $N^{D\pi^-} - N^{D\pi^+}$.  For an isoscalar
target, the contributions from the sea will exactly
cancel, and the result will depend only upon the 
valence quark contributions.  From Eq.\ (\ref{genfor}),
we can show that this quantity can
be written  
\begin{eqnarray}
N^{D\pi^+}(x,z) &-& N^{D\pi^-}(x,z)
  = \nonumber \\ 
  &(& {4\over 9} \left[ u^p_{\rm v}(x) +  u^n_{\rm v}(x)
  \right] - {1\over 9}\left[ d^p_{\rm v}(x)+ 
   d^n_{\rm v}(x)\right] ) 
  \left[ D_u^+(z) - D_u^-(z) \right]
\label{Ndeut}
\end{eqnarray}
In Eq.\ (\ref{Ndeut}), we assume the validity of the impulse approximation, 
i.e., \ $N^{D\pi^+}(x,z) = N^{p\pi^+}(x,z)+ N^{n\pi^+}(x,z)$.  
Note that the valence quark distributions appearing in
Eq.\ (\ref{Ndeut}) are the quark distributions in the
deuteron and {\it not} the free nucleon valence quark 
distributions.   Integrating this quantity over all $x$, 
we obtain  
\begin{eqnarray}
N_D(z) &\equiv& \int_0^1\,dx\,\left[ N^{D\pi^+}(x,z) 
  - N^{D\pi^-}(x,z) \right] \nonumber \\ &=& 
 \left[ D_u^{\pi^+}(z) - D_u^{\pi^-}(z) \right]
\label{dsumrule}
\end{eqnarray}
  
The quantity $N_D(z)$ is proportional to the difference 
between the ``favored''
and ``unfavored'' fragmentation functions, $D_u^{\pi^+}(z)$
and $D_u^{\pi^-}(z)$, respectively.  
If we assume the Feynman--Field parameterization for
the ratio of fragmentation functions, then measurement of
Eq.\ (\ref{dsumrule}) would give
\begin{eqnarray}
N_D(z) &=& {2 z\over 1+z} D_u^{\pi^+}(z) 
\label{Rdz}
\end{eqnarray}
This could then be compared with the fragmentation
functions extracted from the EMC group, or by the 
method suggested by Levelt {\it et al.} \cite{LMS}, which 
requires comparison of pion leptoproduction on both protons 
and neutrons separately.  

The fragmentation functions have previously been 
measured in deep inelastic muon
scattering by the EMC collaboration \cite{emcfrag,emcfrag89}.  
In Fig.\ [1a] we show the favored and unfavored
fragmentation functions for quarks into charged 
pions, as measured by EMC.  In Fig. [1b] we plot the
experimental ratio of fragmentation functions 
$\Delta (z)$, and compare it with the  
Feynman-Field parameterization of Eq.\ (\ref{ffrat}).  
For moderate values of $z$, the Feynman-Field parameterization
is rather accurate.  For the largest values of $z$, the 
experimental errors are large, but the 
experimental results appear to be systematically larger than the
Feynman-Field predictions.  

\subsection{Charge Symmetry Violation in Semi-Inclusive
Pion Production}

We want to measure charge symmetry violation
in the valence quark distributions.  For
pion electroproduction on an isoscalar target, 
(such as the deuteron) this can be achieved by noting that 
(assuming charge symmetry) the `favored' production of
charged pions from valence quarks 
are related by
\begin{eqnarray}
N^{D\pi^+}_{fav}(x,z) &=& 4\,N^{D\pi^-}_{fav}(x,z) .
\end{eqnarray}
That is, for $\pi^+$ ($\pi^-$) production, the ``favored'' mode of
charged pion production is from the target up 
(down) quarks.  Since
the semi-inclusive reactions are proportional to the square
of the quark charge, there is a relative weighting of 4 for 
$\pi^+$ production.  

We therefore propose measuring the quantity 
$R^D(x,z)$, defined by 
\begin{eqnarray}
R^D(x,z) &\equiv&  
{ 4\,N^{D\pi^-}(x,z) - N^{D\pi^+}(x,z)
  \over N^{D\pi^+}(x,z) - N^{D\pi^-}(x,z) } 
  \nonumber \\  &=&  
  {5 \Delta (z)\over 1- \Delta (z)} - {(4+\Delta(z)) \delta D(z) 
  \over 3[1-\Delta(z)]^2} \nonumber \\ 
  &+& \left[ {1+ \Delta (z)\over 
  1- \Delta (z)}\right] \left[ {4\left( \delta d(x) 
  - \delta u(x)\right) 
  + 15\left( \bar{u}^p(x) +  \bar{d}^p(x)\right) \over 
  3\,\left[u^p_{\rm v}(x) + d^p_{\rm v}(x)\right]}\right] 
  \nonumber \\ &+&  
  {\Delta_s(z) \over 1- \Delta (z)} 
  {\left[ s(x) + \bar{s}(x)\right]\over 
  \left[u^p_{\rm v}(x) + d^p_{\rm v}(x)\right]}
\label{sigma}
\end{eqnarray}

The quantity $R^D$ contains the charge symmetry violating
quark distribution functions defined by 
\begin{eqnarray}
\delta d(x) &\equiv& d^p(x)-u^n(x) \nonumber \\ 
  \delta u(x) &\equiv& u^p(x) -d^n(x) ,  \nonumber
\end{eqnarray}
the strange/favored
ratio of quark fragmentation functions
\begin{eqnarray}
\Delta_s(z) &=& {D_s^{\pi^+}(z) + D_s^{\pi^-}(z) \over 
  D_u^{\pi^+}(z)} , \nonumber
\end{eqnarray} 
and the charge symmetry breaking fragmentation functions, 
\begin{eqnarray}
\delta D(z) &\equiv& {D_u^{\pi^+}(z) - D_d^{\pi^-}(z)\over 
  D_u^{\pi^+}(z)}.  
\label{ffcsv} 
\end{eqnarray} 

We can more cleanly separate the $x$ and $z$ 
dependence by multiplying 
$R^D$ by a $z$-dependent factor, e.g.  
\begin{eqnarray}
\widetilde{R}^D(x,z) &\equiv& {1- \Delta (z)\over 
  1+ \Delta (z)}R(x,z) \nonumber \\
  &=& {5 \Delta (z)\over 1+ \Delta (z)} - 
  { (4+\Delta(z))\delta D(z)\over 3\left(1-\Delta^2(z)\right) }  +
  {4\left[ \delta d(x) - \delta u(x)\right]  \over 
  3\,\left[u^p_{\rm v}(x) + d^p_{\rm v}(x)\right]} \nonumber \\ 
  &+& { 5\left( \bar{u}^p(x) +  \bar{d}^p(x) \right) +  
  \Delta_s(z) \left[ s(x) + \bar{s}(x)\right]/(1+ \Delta (z)) 
  \over 
  \left[u^p_{\rm v}(x) + d^p_{\rm v}(x)\right]} \nonumber \\
   &\equiv& \widetilde{R}^D_f(z) + \widetilde{R}^D_{CSV}(x) + 
   \widetilde{R}^D_{sea}(x,z) \quad , 
\label{sigtil}
\end{eqnarray}
where 
\begin{eqnarray}
\widetilde{R}^D_f(z) &=& {5 \Delta (z)\over 1+ \Delta (z)} - 
  { (4+\Delta(z))\delta D(z)\over 3\left( 1-\Delta^2(z)\right) }  
  ; \nonumber \\
  \widetilde{R}^D_{CSV}(x) &=& {4\left[ \delta d(x) - \delta u(x)\right]   
  \over 3\,\left[u^p_{\rm v}(x) + d^p_{\rm v}(x)\right]} ; \nonumber \\ 
  \widetilde{R}^D_{sea}(x,z) &=& { 5\left( \bar{u}^p(x) +  \bar{d}^p(x) 
  \right) + \Delta_s(z) \left[ s(x) + \bar{s}(x)\right]/(1+ \Delta (z)) 
  \over \left[u^p_{\rm v}(x) + d^p_{\rm v}(x)\right]}  \quad . 
\label{sigdef}
\end{eqnarray}
Eq.\ (\ref{sigtil}) is obtained from Eq.\ (\ref{sigma})
by multiplying by the experimentally measured fragmentation
functions.   In Eqs.\ (\ref{sigma}), (\ref{sigtil}) and (\ref{sigdef}), we 
have expanded to first order in
``small'' quantities.  These are: the CSV nucleon terms,
$\delta d(x)$ and $\delta u(x)$; the CSV part of the fragmentation
function, $\delta D(z)$; and the sea quark distributions
(Eq.\ (\ref{sigtil}) is only valid at large $x$ where the
ratio of sea/valence quark distributions is small).  We have
neglected the CSV part of the ``unfavored'' fragmentation function.  
In the region of interest (moderately large $z$) the unfavored
fragmentation function will be considerably smaller than the
favored term, and consequently the unfavored CSV 
term should be proportionately
smaller than the favored CSV term.  

The quantity $\widetilde{R}^D(x,z)$ separates into three pieces.  
The first piece, $\widetilde{R}^D_f(z)$, depends only on $z$, as is 
shown in Eq.\ (\ref{sigdef}).  It contains a small part which is
proportional to the CSV part of the fragmentation function.  The dominant 
piece of $\widetilde{R}^D_f(z)$ has the form 
\begin{eqnarray}
{5 \Delta (z)\over 1+ \Delta (z)} &\approx& 
 {5\,(1-z)\over 2}  , 
\label{delapp}
\end{eqnarray}
where the relation in Eq.\ (\ref{delapp}) follows if
we adopt the Feynman-Field parameterization.  For most values
of $z$, this term should decrease monotonically as $z$ 
increases (although, at the largest values of $z$, it may increase, as
suggested in Fig.\ [1b]).  The second term, $\widetilde{R}^D_{CSV}(x)$ 
depends only on $x$, and is proportional to the nucleon CSV fraction 
(relative to the valence quark distributions).  The term 
$\widetilde{R}^D_{sea}(x,z)$ is proportional to the sea quark 
contributions.  

Experimentally, one needs to measure accurately the  
$x$-dependence of $\widetilde{R}^D(x,z)$ for fixed $z$; 
in this case the $z$-dependent term will be
large (of order one) and constant.  The sea quark 
contribution will be large at small $x$, but should fall
off monotonically and rapidly with $x$.  
So, if one goes to sufficiently large $x$, the sea
quark contribution will be negligible relative to the CSV
term.  One then has to extract the small, $x$-dependent
term in Eq.\ (\ref{sigtil}) from the large term independent 
of $x$.  As a general rule, the larger the values of $x$ 
and $z$ at which data can be taken, the larger the CSV term will
be relative to the $z$-dependent term.  

Note that for a nuclear target, the quark distributions
which appear in Eq.\ (\ref{sigma}) are the {\bf nuclear}
quark distributions, and not the free nucleon distributions.  
However, for the nonstrange quarks a common assumption is
that the quark distributions for flavor $i$, in a nucleus
with $A$ nucleons, are
related to the free ones by $q_i^A(x) = \epsilon(x) q_i^N(x)$.  
If $\epsilon (x)$ is independent of quark
flavor (most models of nuclear effects on quark distributions 
make this assumption), 
then this quantity {\bf cancels} in the ratio 
in Eq.\ (\ref{sigtil}), which would then be identical to 
the ratio for free nucleons.   

\section{Predictions of Charge Symmetry Violation in Pion
Leptoproduction}

\subsection{A Simple Model for the Fragmentation Function into Charged
Pions} 

We want an estimate of the fragmentation function for a quark into
a charged pion.  Our main interest will be to use this simple model
to estimate the magnitude of charge symmetry violation in this
fragmentation function, as this quantity enters into the ratio
$\widetilde{R}^D_f(z)$ of Eq.\ (\ref{sigdef}).  Our calculation is 
based on a method used recently by Nzar and Hoodbhoy \cite{nzar}.  
They calculated the fragmentation function for a quark into a
nucleon plus diquark.  The fragmentation function can be expressed in
terms of the quark field operators and the light cone momentum
fraction $z = P^+/k^+$
\begin{eqnarray}
4{D_{q_i}^{\pi^j}(z) \over z}\,p^+ &=& \sum_X\,\int\,{d\lambda \over 2\pi} 
 e^{-i\lambda/z}
 {\rm Tr} \left\langle 0| \gamma^+ \psi(0)|P;X\right\rangle \,
  \left\langle P;X|\bar{\psi}(\lambda n)|0 \right\rangle
\label{fragdef}
\end{eqnarray}
In Eq.\ (\ref{fragdef}), $D_{q_i}^{\pi^j}(z)$ is the fragmentation 
function for a quark of flavor $i$ and four-momentum $k^\mu$ fragmenting
into a meson of charge $j$ and 
four-momentum $P^\mu$, and the unmeasured state $X$, as is shown 
in Figure 2(a).  We also define null vectors
$p^\mu$ and $n^\mu$ such that $p^2 = n^2 = 0$, $p\cdot n = 1$, and
$n\cdot A = A^+$ for any four vector $A$. 

To calculate the fragmentation function of Eq.\ (\ref{fragdef}), 
we need to evaluate the contribution for every state $X$ and sum 
over the complete set of states.  The simplest approximation is that for
charged pions the fragmentation function is dominated by a 
quark which fragments into a pion with momentum $P^\mu$ and a 
single quark of momentum $(k-P)^\mu$; this is shown schematically
in Figure 2(b).  We further assume that we can approximate 
the amplitude for this process as
\begin{eqnarray}
 \left\langle P;(k-P)|\bar{\psi}|0 \right\rangle &\approx& 
  \bar{u}(k-P)\Phi (k^2){i\over \rlap/k -m }
\label{ampapp}
\end{eqnarray}
In Eq.\ (\ref{ampapp}), a quark with current quark mass $m$ fragments
into a pion with mass $M$ and a quark with constituent mass $m_q$.   
The matrix $\Phi (k^2)$ is given by $\Phi (k^2) = \phi(k^2) \gamma_5$. 

Inserting Eq.\ (\ref{ampapp}) into Eq.\ (\ref{fragdef}), summing
over the intermediate states and using the kinematic conditions
gives 
\begin{eqnarray}
D_q^\pi(z) &=& {z\over 4(1-z)^2}\int_0^\infty \,{dk_\bot^2 \over (2\pi)^2}\,
  {\widetilde{m}_q^2 + z^2k_\bot^2 \over \left( k^2 - m^2 \right)^2 }
  \, |\phi|^2 \qquad , \nonumber \\
\widetilde{m}_q^2 &\equiv& m_q^2 + (1-z)^2 m^2 - 2mm_q(1-z) \qquad .
\label{Dint}
\end{eqnarray}
The kinematic conditions for the fragmenting quark imply that
\begin{eqnarray}
k^2 &=& {M^2\over z} + {m_q^2 + zk_\bot^2 \over 1-z},
\end{eqnarray}
while for the form factor, $\phi$, we choose the damping factor
\begin{eqnarray}
\phi(k^2) = N\,{ k^2 - m^2 \over \left( k^2 - \Lambda^2 \right)^{3/2}}.
\end{eqnarray}
With these assumptions the integral in Eq.\ (\ref{Dint}) can be
done analytically, with the result
\begin{eqnarray}
D_q^\pi(z) &=& {N^2\over 32\pi^2}\, \left( 
  {\widetilde{m}_q^2 \over (1-z)\beta^2} + 
  {z\over \beta} \right) \qquad , \nonumber \\
  \beta &=& {M^2 \over z} + {m_q^2 \over 1-z} - \Lambda^2 \quad .
\label{danal}
\end{eqnarray}

For our calculations we chose $\Lambda = 0.4$ GeV.  For the
fragmentation function $D_u^{\pi^+}$ (i.e., an up quark fragmenting
into a $\pi^+$ and a down quark), we chose $m_u = 5$ MeV and 
$m_q = 353$ MeV.  The resulting theoretical fragmentation functions are
shown as the solid curve in Figure 3(a).  We have arbitrarily 
normalized our fragmentation functions to the experimental 
measurements of the EMC collaboration \cite{emcfrag,emcfrag89}, at
large $z$.  As can be seen, for $z < 0.4$ there is a significant
deviation between the calculated and measured fragmentation functions.
This is not surprising as we expect that the truncation of the complete
set of states, $X$, at a single, constituent quark should only be
reasonable at
large $z$.  

Using these approximations for the fragmentation function, we can
estimate the magnitude of charge symmetry violation in the favored
fragmentation function, by recalculating $D_d^{\pi^-}(z)$ and 
taking into account light quark mass differences.  For this
fragmentation function we used $m_d = 8$ MeV and $m_q = 350$ MeV.
In Figure 3(b) we plot the charge symmetry violating fraction
for the fragmentation function, $\delta D(z)$, given by 
Eq.\ (\ref{ffcsv}).  The predicted CSV term is of the order of a
few percent.  $\delta D(z) < 0$ as expected, since $D_d^{\pi^-}$ 
involves creation of a relatively light $u\bar{u}$ pair compared 
to the heavier $d\bar{d}$ pair created in the case of the fragmentation
function $D_u^{\pi^+}$.  Our simple model predicts a rather large CSV
term.  This is because the pion mass is unusually small,
and the light quark mass differences are small in magnitude but
large as a fraction of the current quark masses.   We emphasize
that this calculation should give only a crude estimate of the
size of CSV terms in the quark fragmentation function.

\subsection{Calculations of CSV 
in Pion Leptoproduction}

In the previous section we proposed measuring the quantity
$\widetilde{R}^D(x,z)$, a ratio of charged pion leptoproduction
cross sections on the deuteron.  As seen from Eqs.\ (\ref{sigtil}) and 
(\ref{sigdef}), 
this quantity separates into three parts: one piece which depends
only on $z$, a second which depends only on $x$ and is proportional
to the quark CSV, and a third part depending on the contribution 
from sea quarks.  

In Fig.\ [4] we plot the $z$-dependent contribution to 
$\widetilde{R}^D(x,z)$, i.e.\ $\widetilde{R}^D_f(z)$ of 
Eqs.\ (\ref{sigtil}) and (\ref{sigdef}).  The solid dots are the
values for this quantity using the EMC fragmentation functions 
\cite{emcfrag,emcfrag89}; they do not include a 
CSV contribution from the fragmentation function, and therefore 
depend only on the ratio of favored to unfavored fragmentation
functions.  The open dots include our
estimate of the charge symmetry violating term from the
fragmentation function, taken from Eqs.\ (\ref{ffcsv}) and 
(\ref{danal}).  The CSV term 
is roughly 1\% of the total $z$-dependent term.  The overall $z$-dependent
term is of order unity; it falls off monotonically and smoothly with
increasing $z$, until $z \approx 0.6$, after which it remains constant
and may increase somewhat.  The solid curve shows this quantity
as approximated by the Feynman-Field parameterization \cite{feyfie}.  

In Fig.\ [5] we plot our predictions for the $x$-dependent
terms in $\widetilde{R}^D(x,z)$, at $Q^2 = 10$ GeV$^2$.  
The long dashed curve is the contribution
from nonstrange sea quarks to $\widetilde{R}^D_{sea}(x,z)$.  
This depends only on $x$, and is calculated using the CTEQ3M parton
distributions from the CTEQ group \cite{cteq}.  The short dashed curve 
is our prediction for the parton charge symmetry violating term, 
$\widetilde{R}^D_{CSV}(x)$; this uses the CTEQ3M parton distributions, 
plus the bag model prediction for valence quark CSV from Londergan 
{\it et al.}
\cite{londergan}.  The dot-dashed curve is the contribution from
strange quarks.  This last term is proportional to the strange quark
fragmentation function.  As we do not know this, we have removed 
the $z$ dependence of this term by approximating $\Delta_s(z)/\left(
1+\Delta(z) \right) \approx 1$.  This will substantially overestimate 
the strange quark contribution, but even with this approximation the 
strange quarks contribute only a small fraction of the total.  The solid 
curve is the sum of the three terms.  

\,From Fig.\ [5] we see that for $x \approx 0.5$, the CSV term is as large as
the sea quark contribution, and with increasing $x$ the 
CSV term dominates the $x$ dependence of this ratio.  We
predict the maximum CSV contribution will be of order 
$0.02 - 0.04$.  In the small-$x$ region
dominated by the sea quarks, $\widetilde{R}^D_{sea}(x,z)$ (at
constant $z$) should decrease rapidly with increasing $x$.  
However, for $x \geq 0.55$ the CSV term is predicted to dominate this 
ratio, and at the largest values of $x$ the terms 
$\widetilde{R}^D_{CSV}(x)+ \widetilde{R}^D_{sea}(x,z)$ 
should be dominated by the CSV term.  

If we measure the $x$ dependence of $\widetilde{R}^D(x,z)$ at
constant $z$, then the $x$ dependent contribution shown in
Fig.\ [5] will sit on a large and constant $z$ dependent term,
estimated in Fig.\ [4].  At small $x$ the sea quark term should
be rapidly and smoothly falling with $x$.  The value of the 
constant $z$-dependent background 
can be obtained by extrapolating the sea quark contribution
to zero at large $x$; the CSV contribution would then be 
the difference between this extrapolated value and the measured
value at large $x$.   

From our previous work on parton charge symmetry 
violation \cite{Rodionov,londergan} we predict that the two CSV
terms in Eq.\ (\ref{sigtil}) 
(i.e., $\delta d(x)$ and $\delta u(x)$) 
will be roughly equal in absolute value (each of them 
should be of the order 1-2\% of the average up + down valence
quark distribution), and they should have opposite signs.  As
a result, in Eq.\ (\ref{sigtil}) the two CSV contributions
would add constructively, and should produce a term whose
value is of the order of $0.02-0.04$.  Sather's CSV 
parton distributions \cite{Sather} have the same qualitative behavior 
as ours.  

We predict that the $z$-dependent term will be  
much larger then the $x$-dependent terms, as can be seen from 
comparing Figs.\ [4] and [5].  Using the EMC measured values for
the fragmentation functions, as $z$ 
goes from 0.4 to 0.8 the $z$-dependent term in Eq.\ (\ref{sigtil})
should vary between 
approximately 1.5 and 1.  So the CSV term is expected
to be between 1-4\% of the $z$-dependent term.  

One major result of our calculations is the prediction that the 
$x$ and $z$-dependent parts of $\widetilde{R}^D(x,z)$ will separate.  
Our current calculation has 
been carried out in the ``naive parton model;'' we have not included 
things like 
higher-order contributions to leptoproduction or scaling
violations.
Since we 
predict that CSV terms will contribute at the few percent level, 
it will be necessary to investigate whether
these higher-order contributions 
to charged pion electroproduction are negligible compared with 
our lowest-order contributions.  Another question
is whether these additional contributions will preserve the 
separation of variables $x$ and $z$ in 
the quantity $\widetilde{R}^D(x,z)$ of Eq.\ (\ref{sigtil}).  These will
require more sophisticated calculations, which we are presently
undertaking.  

In the HERMES experiment at HERA, the goal is to make 
precision measurements of the spin structure functions.  For
this one must know very accurately the spin dependence of high
energy electron scattering from deuterium.  This requires 
precise knowledge of the sources
of systematic error, so the prospect for obtaining very
accurate spin-averaged charged pion leptoproduction data is
excellent.  Only data from deuterium targets is required; efficient
detection of both signs of charged pions is important, but absolute
yields are not required as overall normalizations cancel out in
the ratio of Eq.\ (\ref{sigtil}). 

\section{Conclusions and Future Experiments}

In conclusion, we have shown that one can use charged
pion electroproduction on an unpolarized deuteron 
target to measure 
the charge symmetry violating [CSV] contribution in
the nucleon's valence parton distributions.  
These are expected to be of the order of a few
percent, so one has to extract
a small term in the presence of a large background.   
Since one of the terms depends only on $z$, and the other
only on $x$, one expects to exploit this feature 
experimentally. This would constitute the first direct
measurement of charge symmetry violation in the valence quark
distribution functions.      

In defining parton distribution functions, it is routinely assumed 
that charge symmetry is valid.  Surprisingly
enough, there are very few tests of this assumption.  A reasonable
estimate of charge symmetry violation would be at roughly the 1\% level,
but it is conceivable that charge symmetry violation could be a
few times this value and would not have been observed to date.  
An experimental upper limit for charge symmetry in valence quark
distributions would be very useful.  Measurement of a nonzero
effect would be extremely interesting.  In this paper we have
explained why the measurement of charged
pion leptoproduction from an isoscalar target, for example
charged pion electroproduction from deuterium, would constitute an
excellent test of charge symmetry in the valence
quark parton distributions. We also estimated the charge symmetry
violation in the pion fragmentation functions for the first time and
showed that it would not interfere with the extraction of information
about the nucleon valence distributions. 
As measurements of semi-inclusive charged pion production on the
deuteron are currently in progress
in the HERMES experiment at HERA, we are hopeful that suitable data might
soon be available.  

This work was supported by the Australian Research Council.  
Two of the authors [JTL and AP] were supported in part
by the US NSF under research contract NSF-PHY94-08843.  
The authors wish to thank Mr.\ S. Braendler
for assistance in calculating parton distributions, and H. Jackson
for discussions regarding the HERMES collaboration measurements.  Two 
of the authors [JTL and AP] wish to thank A. Szczepaniak for useful
discussions.  One of the authors [JTL] wishes to thank the Dept. of
Physics and Mathematical Physics, and the Institute 
for Theoretical Physics, of the University of Adelaide, 
for their hospitality during a visit where this work
was undertaken.

\newpage
{\Large FIGURE CAPTIONS}
\vspace*{0.5cm}
\normalsize

{\bf 1.}  Fragmentation function for quarks into pions, vs.\ 
momentum fraction $z$.  (a) Circles: favored fragmentation function 
$D_u^{\pi^+}(z)$; triangles: unfavored fragmentation 
$D_u^{\pi^-}(z)$, vs.\ $z$.  Data is from the EMC collaboration, 
Ref.\ \cite{emcfrag} for open points and Ref.\ \cite{emcfrag89} for
solid points.  (b) Unfavored/favored ratio 
$\Delta(z) = D_u^{\pi^-}(z)/D_u^{\pi^+}(z)$.  Data from the EMC
collaboration.  Solid curve: theoretical prediction of Feynman and
Field, Ref.\ \cite{feyfie}.  

{\bf 2.}  (a) Fragmentation of quark with momentum $k$ into a
pion with momentum $P$, and any remaining state $X$.  (b) 
Approximation to fragmentation function where the fragmentation
is assumed to be dominated by a quark-pion-quark vertex. 

{\bf 3.}  (a) Estimate of quark-pion favored fragmentation function
$D_u^{\pi^+}(z)$.  Solid curve: model result given by Eq.\ (\ref{danal}), 
assuming $m_u = 5$ MeV, $m_q = 353$ MeV, $\Lambda = 400$ MeV; 
experimental points are those of the EMC group, Ref.\ \cite{emcfrag,emcfrag89}.  
Dashed curve: fragmentation function $D_d^{\pi^-}(z)$, from 
Eq.\ (\ref{danal}), with $m_d = 8$ MeV, $m_q = 350$ MeV.
(b) Estimate of charge symmetry violation in fragmentation function.  
The quantity $\delta D(z)$ of Eq.\ (\ref{ffcsv}) and the model 
fragmentation functions of Fig.\ (3a).  

{\bf 4.}  The quantity $\widetilde{R}^D_f(z)$
of Eq.\ (\ref{sigdef}).  Fragmentation functions taken from the
EMC experiment, Ref.\ \cite{emcfrag,emcfrag89}.  
Solid dots: does not include 
charge symmetry violation in fragmentation function; open dots: 
includes estimate of CSV in fragmentation function, from Eqs.\ 
(\ref{ffcsv}) and \ref{danal} (and shown in Fig.\ [3]).  Solid curve:
the quantity $\widetilde{R}^D_f(z)$ using the Feynman-Field 
approximation to the fragmentation functions, Ref.\ \cite{feyfie}.  

{\bf 5.}  $x$-dependent contributions to $\widetilde{R}^D(x,z)$
of Eq.\ (\ref{sigtil}).  Long-dashed curve: contribution from nonstrange 
sea quarks; dashed curve: $\widetilde{R}^D_{CSV}(x)$
of Eq.\ (\ref{sigdef}), the contribution from CSV in nucleon parton
distributions; dot-dashed curve: estimate of contribution to 
$\widetilde{R}^D_{sea}(x,z)$ from strange quarks; 
solid curve: total contribution including both
sea quark and CSV contributions.  Parton distributions (CTEQ3M)
are from the CTEQ group, Ref.\ \cite{cteq}, at $Q^2 = 10$ GeV$^2$; CSV 
term is from bag model calculation of Londergan {\it et al.}, Ref.\ 
\cite{londergan}.


\newpage
\begin{figure}
\noindent
\epsfysize=3.8in
\hspace{1.8in}
\epsffile{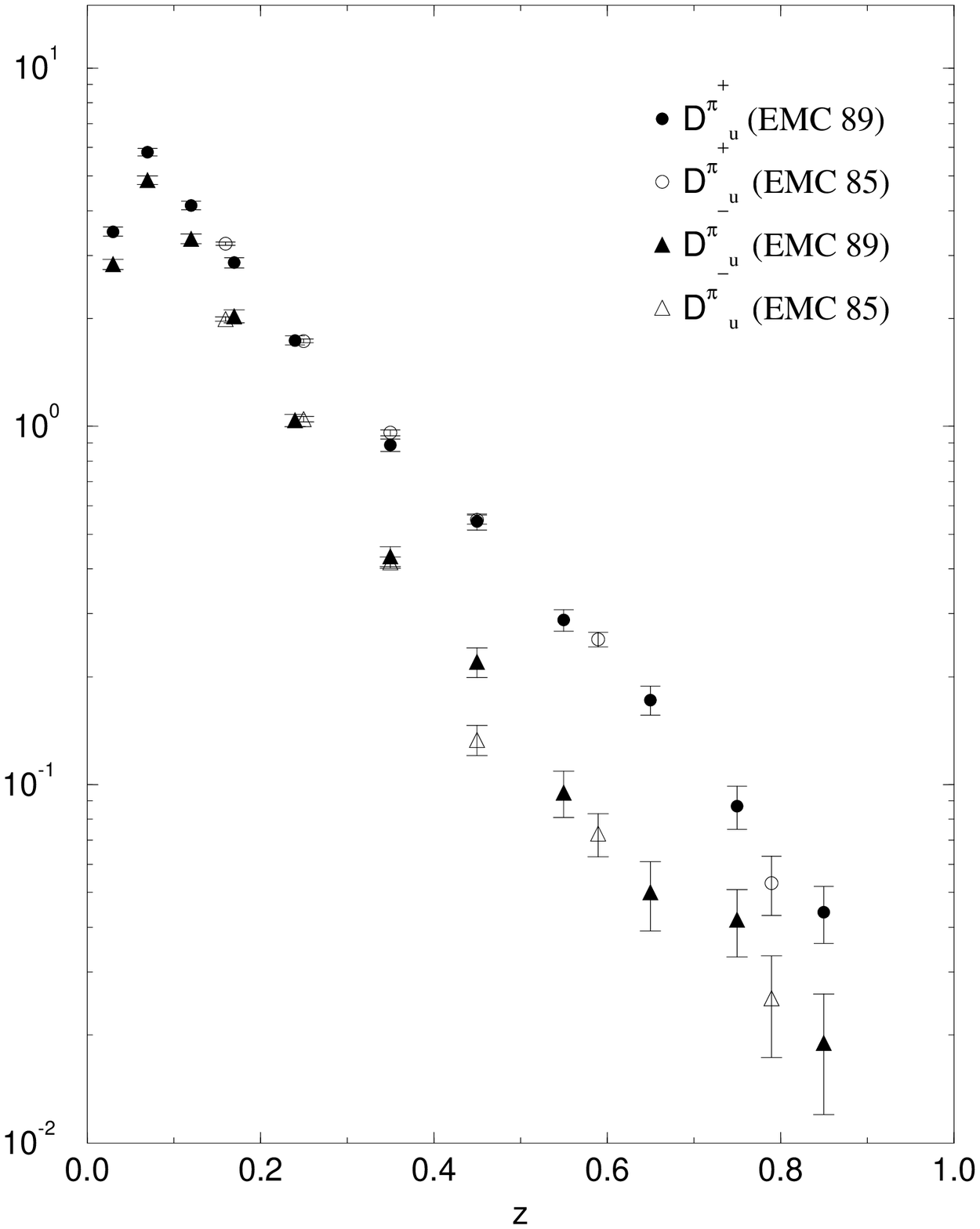}
\begin{center} Fig. 1 (a) \end{center}
\epsfysize=3.8in
\hspace{1.8in}
\epsffile{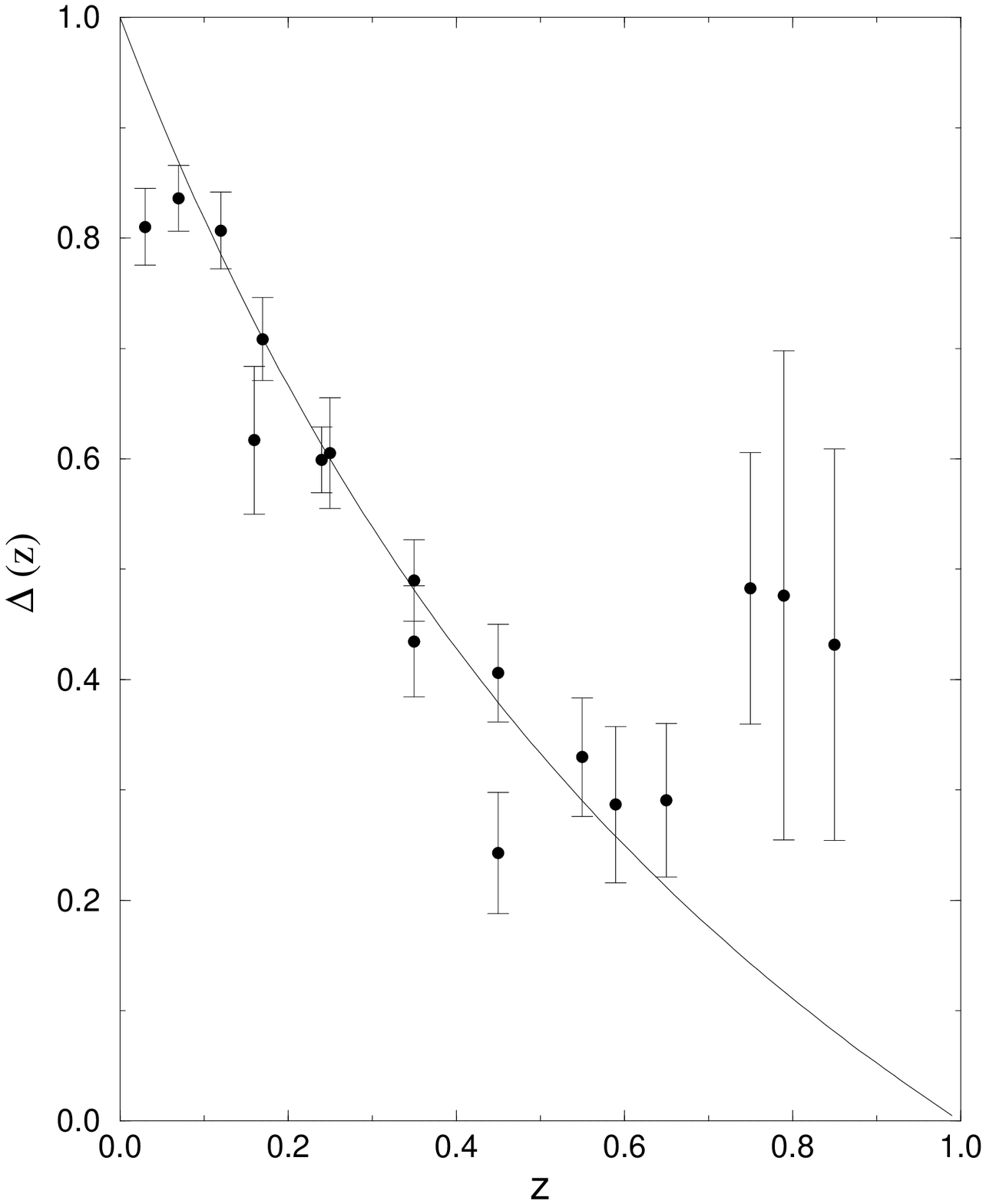}
\begin{center} Fig. 1 (b) \end{center}

\newpage
\epsfxsize=3.5in
\hspace{1.2in}
\epsffile{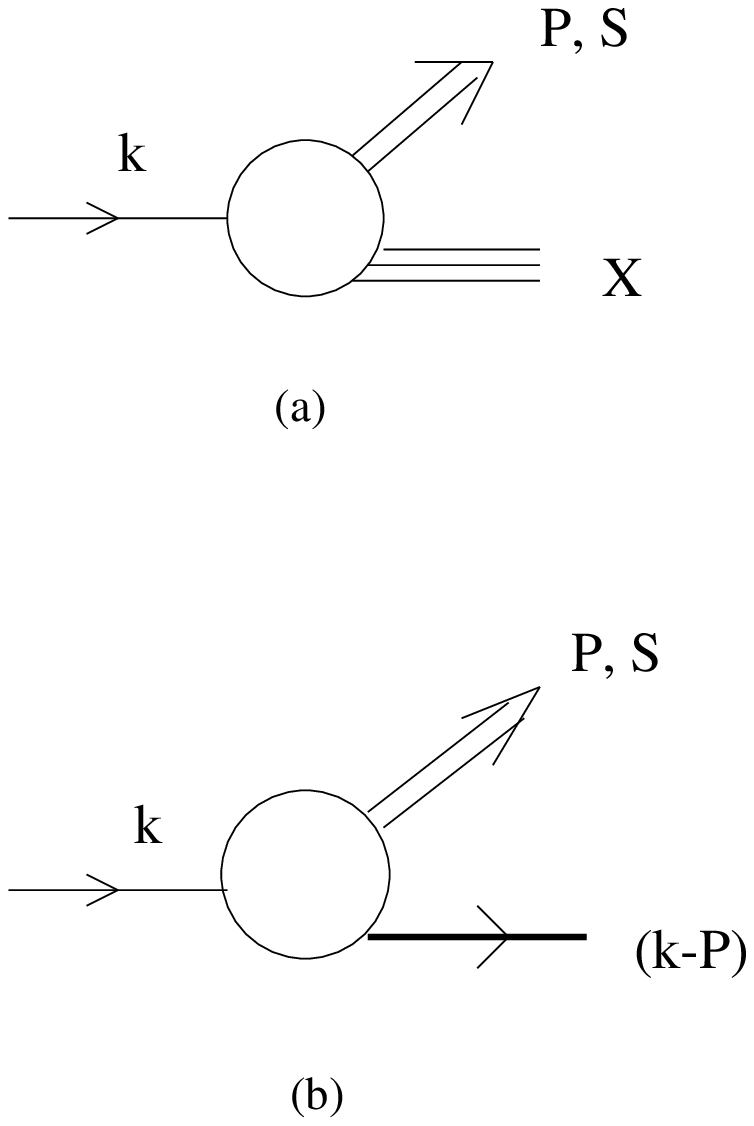}
\vspace{1.in}
\begin{center} Fig. 2 \end{center}

\newpage
\noindent
\epsfysize=4.in
\hspace{1.8in}
\epsffile{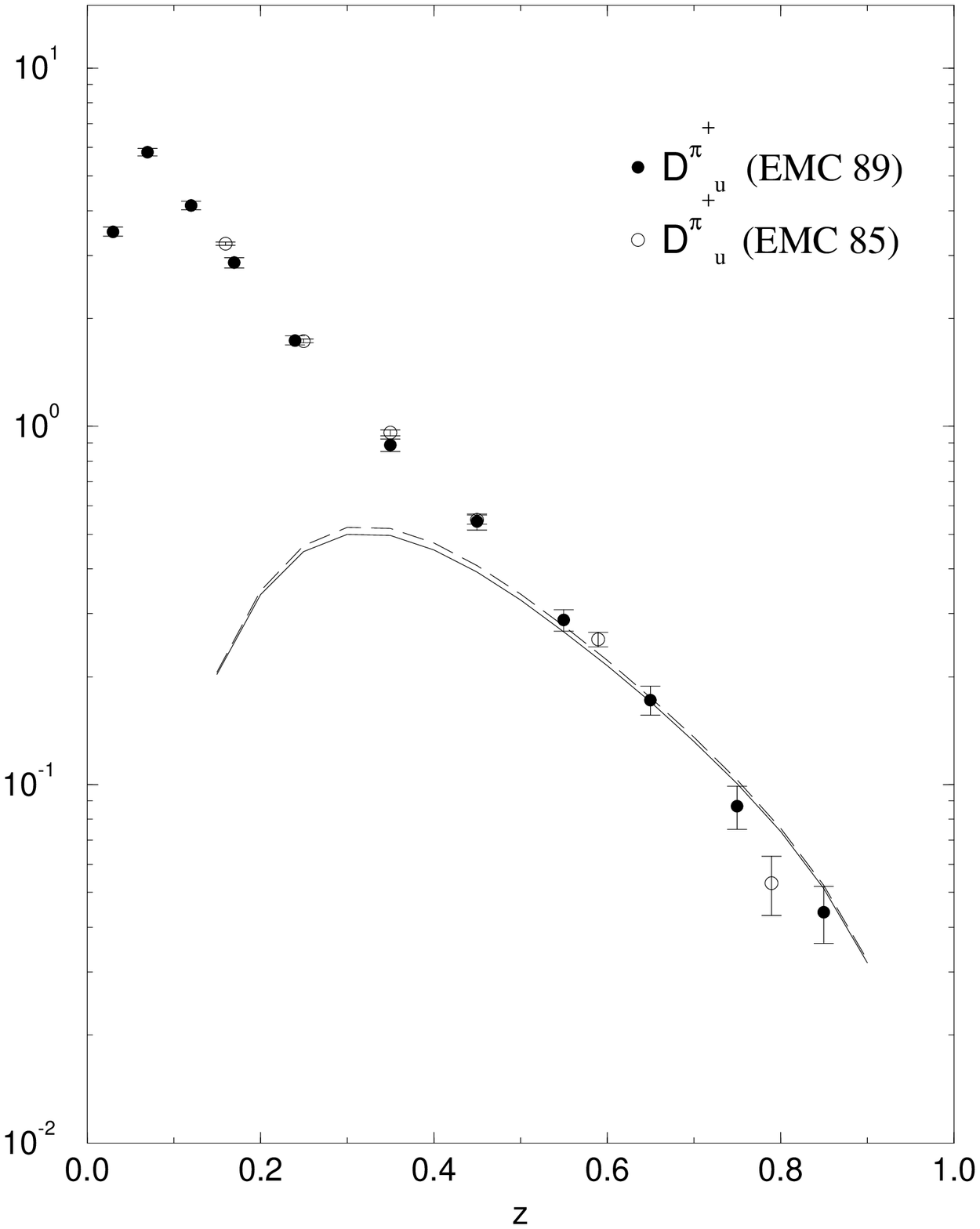}
\begin{center} Fig. 3 (a) \end{center}
\epsfysize=4.in
\hspace{1.8in}
\epsffile{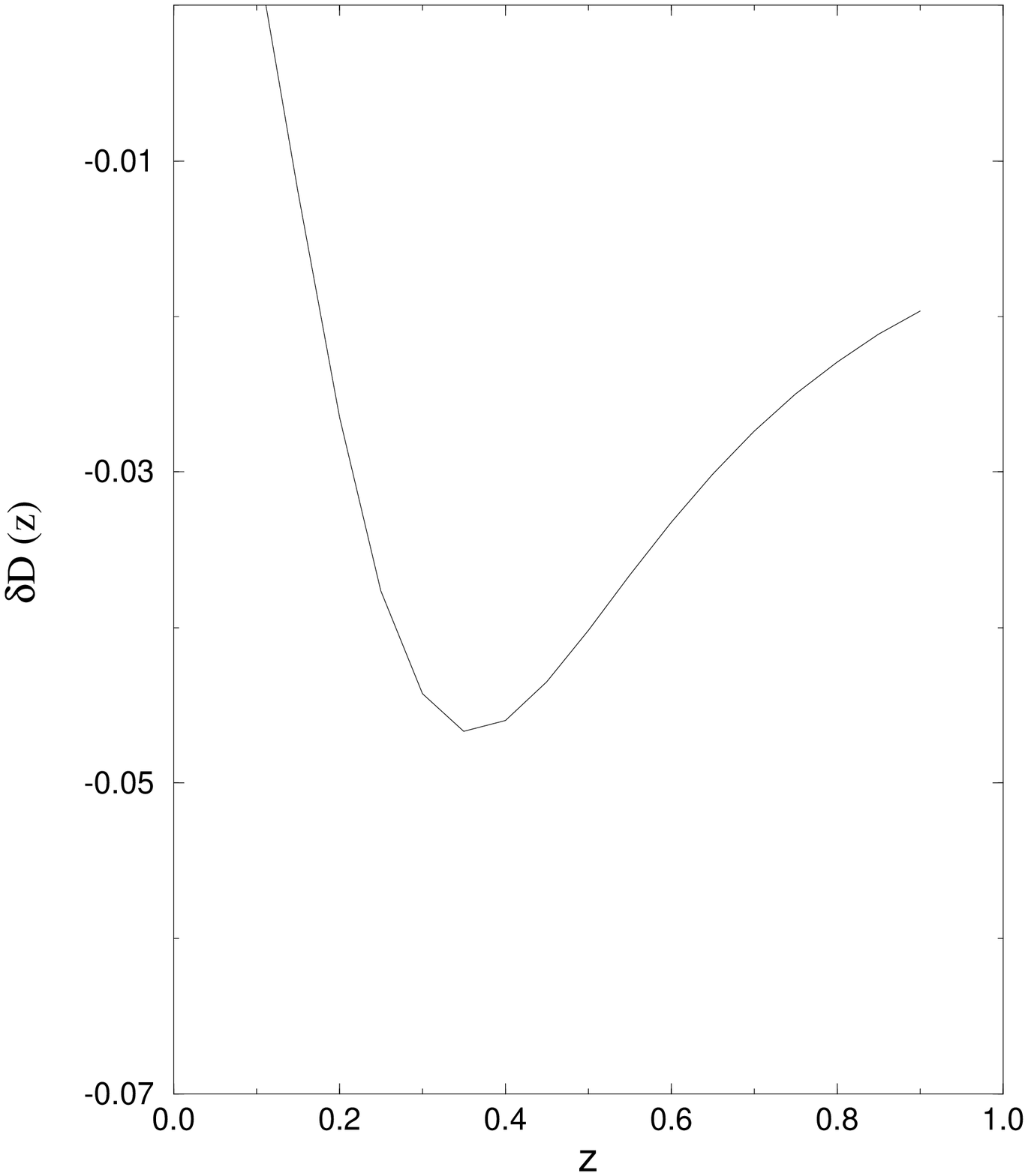}
\begin{center} Fig. 3 (b) \end{center}

\newpage
\epsfysize=6.5in
\hspace{1.2in}
\epsffile{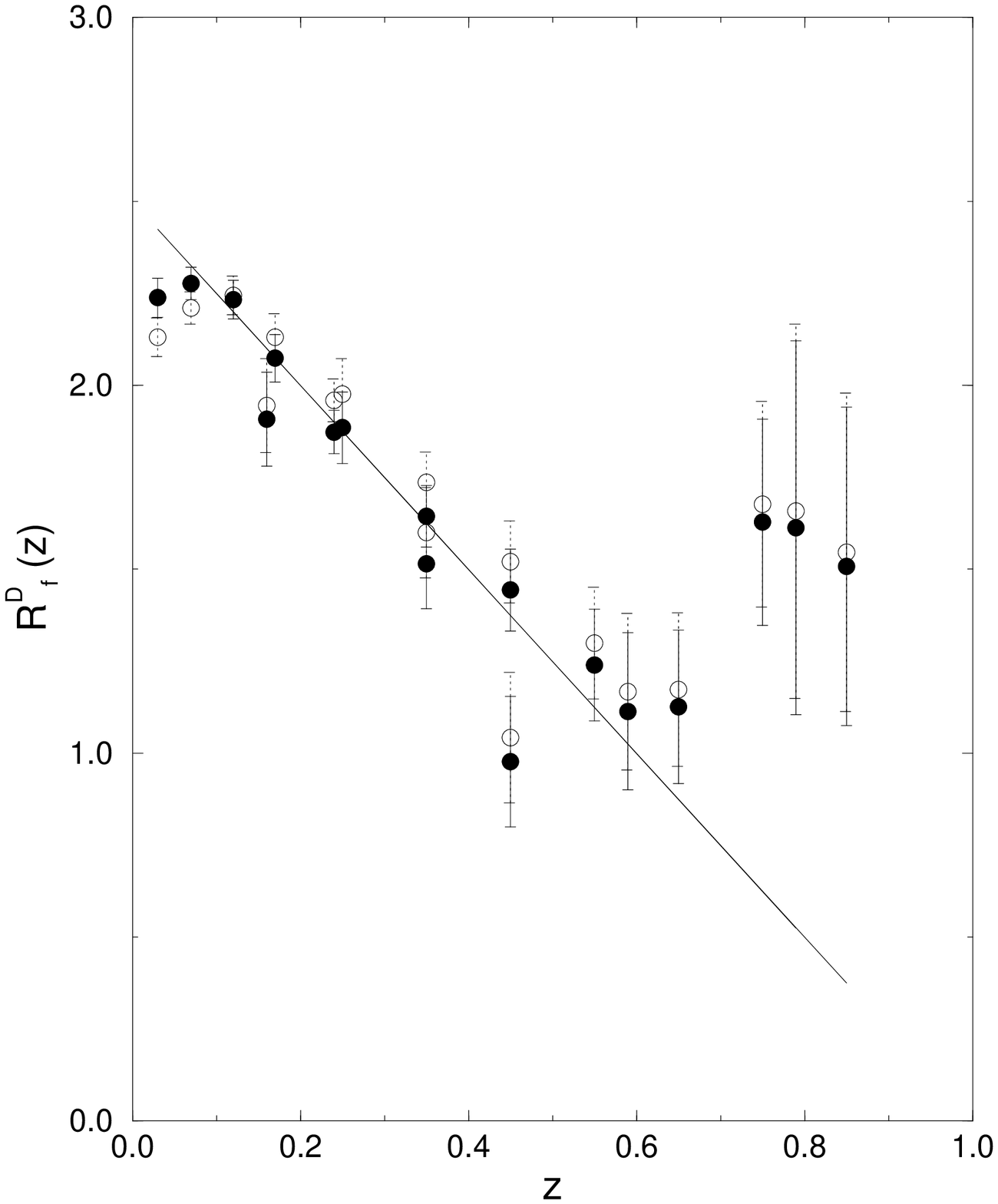}
\begin{center} Fig. 4 \end{center}

\newpage
\epsfysize=6.5in
\hspace{1.2in}
\epsffile{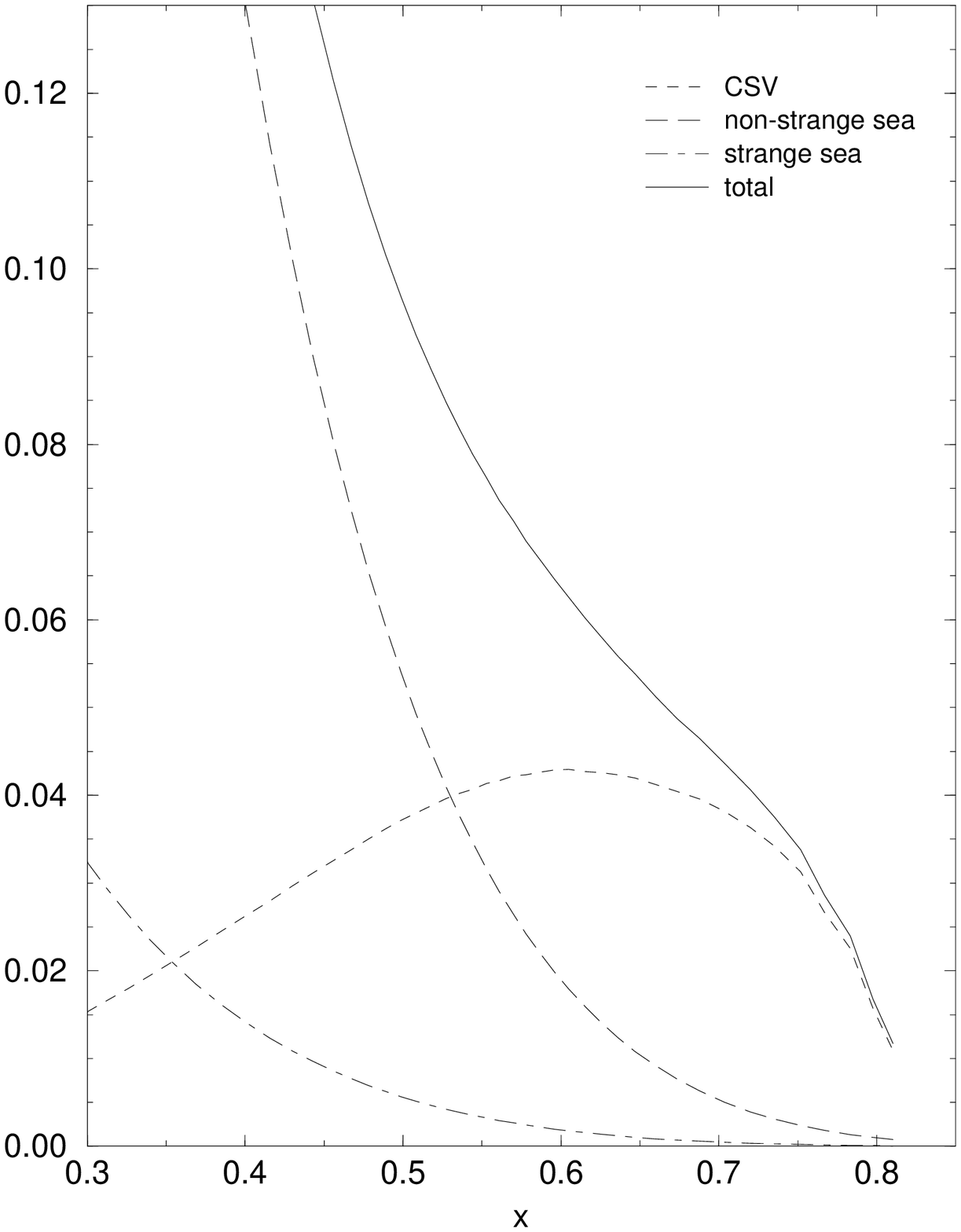}
\begin{center} Fig. 5 \end{center}

\end{figure}

\end{document}